\documentclass[twocolumn,prb,showpacs,preprintnumbers]{revtex4}

\usepackage{graphicx}
\usepackage{dcolumn}
\usepackage{bm}
\usepackage{epsfig}
\usepackage{multirow}

\begin{document}


\title{Bulk and shear relaxation in glasses and highly viscous liquids}

\author{U. Buchenau}
\email{buchenau-juelich@t-online.de}
\affiliation{%
Institut f\"ur Festk\"orperforschung, Forschungszentrum J\"ulich\\
Postfach 1913, D--52425 J\"ulich, Federal Republic of Germany}%
\date{April 2, 2012}

\begin{abstract}
The ratio $\delta B/\delta G$ between the couplings of a relaxational process to compression and shear, respectively, is calculated in the Eshelby picture of structural rearrangements within a surrounding elastic matrix, assuming a constant density of stable structures in distortion space. The result is compared to experimental data for the low-temperature tunneling states in glasses and to Prigogine-Defay data at the glass transition, both from the literature.
\end{abstract}


\maketitle

\section{Introduction}

An amorphous solid reacts to an instantaneous affine shear deformation with the infinite frequency shear modulus $G_\infty$. But the atoms are not in equilibrium then; there is a non-affine motion of the atoms which lowers the energy \cite{wittmer,leonforte}, leading to a lower high-frequency modulus $G$. This motion is intimately related to the boson peak and to the tunneling states which dominate the glass behavior at very low temperatures \cite{bggprs,schirmacher} as well as to the plastic modes responsible for the shear thinning in Non-Newtonian flow \cite{procaccia}. $G$ is the shear modulus which one measures in a light-scattering Brillouin experiment at GHz frequency, a frequency markedly lower than the one of the boson peak.

In undercooled liquids, the flow process becomes so slow that one begins to see the shear modulus $G$ for high enough frequencies \cite{yoshino,furukawa}. According to a very recent replica (or cloned liquid) theory result \cite{yoshino}, the point in temperature where one begins to see the shear modulus is the critical temperature of the mode coupling theory \cite{gotze}, though numerical simulations \cite{harrowell} see it already at a higher temperature.

If one begins to see a high frequency shear modulus, one also has to distinguish between the high-frequency and zero frequency bulk moduli $B$ and $B_0$, respectively ($B_0$ is the inverse of the isothermal compressibility). Though there is a standard textbook description of the viscoelastic liquid \cite{ferry}, there is no theory relating $B-B_0$ to $G$. 
The mode coupling theory does not contain a shear modulus; in fact, if the new cloned liquid theory result \cite{yoshino} is correct, the theory is no longer valid at the temperature where the shear modulus appears. The random first order theory (RFOT) \cite{wolynes} is supposed to be valid, but the RFOT specifies only the entropy and the surface tension of a rearranging region and thus does not supply the needed information. Similarly, the coupling model \cite{ngai} does not specify a difference between compression and shear. The elastic models \cite{nemilove,dyre} postulate a proportionality of the flow barrier to $G$ (one wonders a bit whether they should not propose a proportionality to $G_\infty$), but they make no selective prediction for compression or shear. The counting of covalent constraints \cite{jcphillips,thorpe} allows one to count the number of floppy modes per atom in a given substance, but tells nothing about a compression-shear ratio.

There are some specific atomic motion models, in particular for close-packed atoms interacting with a simple two-body potential, which do allow a prediction. One of them is the interstitial model \cite{granato}. In the crystal, the interstitial does not couple to an external compression, but only to an external shear \cite{inter}. Another one is the gliding triangle model \cite{buscho}, which postulates a shear motion of two adjacent triangles of atoms with respect to each other. Both models predict the compression-shear ratio zero, because there is only a coupling to the shear. This result is indeed compatible with experiment and simulation in Lennard-Jones and metallic glasses \cite{pri}, but not in other substances.

The only theoretical concept able to derive a prediction on general grounds is the Eshelby picture of local structural rearrangements within an elastic solid \cite{eshelby,mura,asymm,ascom}. The present paper derives a general relation for the compression-shear ratio on this basis. 

The paper proceeds by the derivation of the general relation in Section II. Section III compares to experimental data from glasses and liquids. Section IV concludes and summarizes the paper.

\section{Theoretical basis}

\subsection{Bulk and shear relaxation in the Eshelby picture}

\begin{figure}
\hspace{-0cm} \vspace{0cm} \epsfig{file=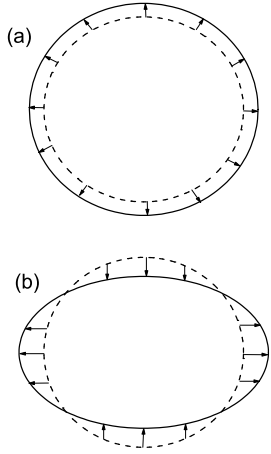,width=4cm,angle=0} \vspace{0cm}\caption{The two types of Eshelby rearrangements (a) a pure volume change $u_i=\delta{\rm v}_i/{\rm v}$ (b) a pure shear $e_i$.}
\end{figure}

In the Eshelby scheme, one considers a strain-free ground state of a volume $N{\rm v}_a$ of $N$ atoms or molecules with an atomic or molecular volume ${\rm v}_a$, embedded in an elastic matrix with the same elastic constants as the volume itself. A structural rearrangement to an alternative stable structure of the volume changes its shape. To first order, the shape change is a pure distortion, describable in terms of the inner volume change $u_i=\delta{\rm v_i}/{N\rm v_a}$ and five linearly independent shear angles $e_i$ (see Fig. 1). 

After the rearrangement, both the inner and the outer volume end up in a strained state, because the alternative stable state has a misfit with respect to the matrix. The resulting elastic strain energy $\Delta$ is about half the energy needed to squeeze the new stable state into the old hole without deforming it. 

Let the initial volume be a sphere. According to Eshelby \cite{eshelby,mura}, for a specific pure volume change $u_i$
\begin{equation}\label{deltab}
	\Delta_b=\frac{2B G}{3B+4G}N{\rm v_a}u_i^2=\frac{2}{9}\frac{1+\nu}{1-\nu}G N{\rm v_a}u_i^2,
\end{equation}
where the index $b$ stands for a bulk distortion and $\nu$ is the Poisson ratio, related to the ratio $B/G$ by
\begin{equation}\label{poisson}
	\frac{B}{G}=\frac{2(1+\nu)}{3(1-2\nu)}.
\end{equation}
The coupling constant $\gamma_b$ to an external volume change $u=\delta{\rm v}/{\rm v}$ is given by
\begin{equation}\label{gammab}
	\gamma_b=-\frac{\partial\Delta_b}{\partial u_i}=-\frac{4}{9}\frac{1+\nu}{1-\nu}G N{\rm v_a}u_i.
\end{equation}

Similarly \cite{eshelby,mura}, for a pure shear change $e_i$
\begin{equation}\label{deltas}
	\Delta_s=\frac{7-5\nu}{30(1-\nu)}G N{\rm v_a}e_i^2,
\end{equation}
where the index $s$ stands for a shear distortion. The coupling constant $\gamma_s$ to an external shear $e$ in the direction $e_i$ is
\begin{equation}\label{gammas}
	\gamma_s=-\frac{\partial\Delta_s}{\partial e_i}=-\frac{7-5\nu}{15(1-\nu)}G N{\rm v_a}e_i.
\end{equation}

For equal asymmetries $\Delta_b=\Delta_s$ one has
\begin{equation}\label{gbgs}
	\frac{\delta B}{\delta G}=\frac{\gamma_b^2}{\gamma_s^2}=\frac{20}{3}\ \frac{1+\nu}{7-5\nu}.
\end{equation}

In order to set the volume change on the same energy scale as the shear distortion, one defines the variable $e_6$ with
\begin{equation}\label{e6}
	e_6^2=\frac{20}{3}\ \frac{1+\nu}{7-5\nu}u_i^2.
\end{equation}

\begin{table*}[htbp]
	\centering
		\begin{tabular}{|c|c|c|c|c|c|c|c|c|c|c|c|c|c|}
\hline
subst.&$\overline{M}$&$T_g$&$\rho$&$\alpha_g$&$\alpha_l$&$B$&$B_0$&$c_{pg}$&$c_{pl}$&$\nu$&$\Pi$&\multicolumn{2}{|c|}{$\delta B/\delta G$}          \\
\hline   
 &a. u.& K &kg/m$^3$&\multicolumn{2}{|c|}{10$^{-4}$K$^{-1}$}&\multicolumn{2}{|c|}{GPa}&\multicolumn{2}{|c|}{10$^6$J/m$^3$}& &  &at T$_g$&tunn. \\ 
\hline
a) silicates: &&&&&&&&&&&&& \\  
SiO$_2$&20.00&1480&2198&.017&.009&45.05&12.66&2.68&2.98&.18&$>$100&1.23&1.2$^a$ \\
15.4 \% Na$_2$O&20.10&773&2314&.23&.55&36.10&15.87&2.66&3.01&.24&15.19&1.53& \\
20.5 \% Na$_2$O&20.14&746&2352&.32&.74&35.09&17.06&2.74&3.20&.27&10.66&1.68& \\
25 \% Na$_2$O&20.17&736&2380&.40&.95&35.21&16.67&2.81&3.36&.29&7.62&1.80&  \\
25.5 \% Na$_2$O&20.17&734&2392&.40&.97&35.21&18.52&2.82&3.37&.29&5.90&1.69&  \\
26.6 \% Na$_2$O&20.18&733&2403&.42&1.02&35.21&18.52&2.84&3.42&.30&5.70&1.72&  \\
29.7 \% Na$_2$O&20.20&725&2415&.47&1.19&34.97&19.31&2.87&3.53&.31&4.03&1.63&\\
33.4 \% Na$_2$O&20.22&715&2433&.53&1.42&35.09&20.51&2.92&3.65&.33&2.61&1.36&\\
39.6 \% Na$_2$O&20.26&683&2457&.63&1.83&34.97&21.01&2.97&3.81&.35&1.61&.89&\\
43.0 \% Na$_2$O&20.28&671&2463&.69&2.06&35.09&20.62&3.01&3.89&.37&1.40&.76&\\
49.5 \% Na$_2$O&20.33&648&2475&.78&2.56&35.21&20.12&3.07&4.03&.40&1.00&.01&\\
12.4 \% K$_2$O&21.40&795&2293&.23&.64&32.57&14.10&2.50&2.94&.23&12.93&1.45&\\
19.5 \% K$_2$O&22.22&763&2347&.33&.95&30.67&15.31&2.42&2.93&.27&5.75&1.49&\\
23.6 \% K$_2$O&22.69&748&2375&.39&1.11&29.76&14.84&2.38&2.97&.28&5.18&1.60&\\
28.4 \% K$_2$O&23.24&729&2398&.46&1.30&29.15&14.97&2.40&3.00&.31&3.75&1.61&\\
33.9 \% K$_2$O&23.86&713&2421&.53&1.52&28.90&16.21&2.42&3.03&.33&2.35&1.33&\\
39.0 \% K$_2$O&24.45&683&2427&.59&1.74&27.86&17.48&2.39&3.01&.35&1.47&.66&\\
\hline
b) borates: &&&&&&&&&&&&& \\  
B$_2$O$_3$&13.92&550&1792&.54&4.00&10.00&2.56&2.37&3.44&.28&4.71&1.77&1.66$^b$\\
6.2 \% Na$_2$O&14.34&605&1949&.31&3.92&19.61&2.93&2.77&4.00&.29&4.52&1.93&\\
9.3 \% Na$_2$O&14.55&643&2000&.28&3.35&22.27&3.95&2.96&4.18&.29&4.19&1.94&\\
11.1 \% Na$_2$O&14.67&663&2040&.27&3.05&25.51&4.55&3.07&4.35&.29&4.49&1.97&\\
14.1 \% Na$_2$O&14.87&707&2079&.26&2.45&27.25&5.41&3.22&4.43&.30&5.29&2.03&\\
15.7 \% Na$_2$O&14.98&726&2105&.26&2.30&27.55&5.99&3.26&4.48&.30&5.31&2.05&\\
17.5 \% Na$_2$O&15.10&731&2132&.27&2.15&27.70&6.69&3.26&4.54&.30&5.62&2.06&\\
19.2 \% Na$_2$O&15.20&736&2155&.28&2.10&29.41&7.57&3.25&4.59&.31&5.38&2.07&\\
21.4 \% Na$_2$O&15.36&741&2188&.28&2.07&31.85&8.36&3.19&4.57&.31&5.12&2.09&\\
24.6 \% Na$_2$O&15.57&743&2227&.31&2.26&35.97&10.78&3.16&4.57&.31&3.24&1.95&\\
29.8 \% Na$_2$O&15.93&743&2304&.38&3.00&40.00&10.60&3.36&4.86&.32&2.04&1.80&\\
35.0 \% Na$_2$O&16.28&723&2325&.41&3.82&40.82&9.42&3.07&4.91&.33&1.79&1.84&\\
4 \% K$_2$O&14.34&597&1923&.46&4.05&18.87&2.42&2.81&3.98&.28&5.49&1.92&\\
5 \% K$_2$O&14.44&603&1937&.45&3.95&19.61&2.75&2.87&4.05&.28&5.00&1.91&\\
8 \% K$_2$O&14.76&627&1988&.40&3.45&15.55&3.56&3.24&4.37&.29&4.21&1.84&\\
10 \% K$_2$O&14.96&643&2016&.38&3.15&22.68&4.74&3.21&4.52&.29&4.43&1.89&\\
15.5 \% K$_2$O&15.54&677&2066&.35&2.35&25.06&6.01&3.37&4.67&.29&6.08&1.97&\\
19.5 \% K$_2$O&15.96&697&2092&.35&2.10&25.97&6.85&3.33&4.54&.30&6.11&2.00&\\
20 \% K$_2$O&16.02&698&2096&.35&2.10&25.97&7.20&3.33&4.55&.30&5.71&1.99&\\
30 \% K$_2$O&17.07&725&2212&.45&3.00&28.82&8.35&2.88&3.89&.31&1.84&1.53&\\
\hline		
		\end{tabular}
	\caption{Coupling ratio $\delta B/\delta G$ from Prigogine-Defay data and Poisson ratios for silicates and borates at the glass transition. Prigogine-Defay data (average atomic mass $\overline{M}$, glass temperature $T_g$, density $\rho$, glass and liquid thermal volume expansion coefficients $\alpha_g$ and $\alpha_l$, glass and liquid bulk moduli $B$ and $B_0$, glass and liquid heat capacities per unit volume at constant pressure $c_{pg}$ and $c_{pl}$) taken from reference \cite{nemilov}. The Poisson ratio for silica is from Bucaro and Dardy \cite{bucaro}, the one for 33\% Na$_2$O from Webb \cite{webb}. For the other silicates, the Poisson ratios were interpolated assuming a linear dependence on concentration with equal coefficients for Na$_2$O and K$_2$O. The Poisson ratio for the borates taken (again assuming a linear concentration dependence) from reference \cite{dangelo}. Tunneling data $^a$Berret and Meissner \cite{berret}; $^b$Pohl et al \cite{pohl}.}
	\label{tab:rse}
\end{table*}

For the surface of the sphere $\Delta$=const in the six-dimensional distortion space $e_1..e_6$, one finds the maximum value for $\gamma_b^2$ along the $e_6$-axis and the maximum value for $\gamma_s^2$ along any of the five shear axes. These maximum values have again the ratio of eq. (\ref{gbgs}). For a constant density of stable states in distortion space, the average values are just 1/6 of these maximum values. Therefore eq. (\ref{gbgs}) should also hold for the average values on the sphere surface. Integrating over $\Delta$, one then gets the relation for the whole ensemble.

\subsection{The thermodynamics of the glass transition}

At the glass transition, $\delta B/\delta G$ cannot be directly determined from $(B-B_0)/G$, because one has to take the enthalpy-volume correlation into account \cite{gundermann,pri}. At $T_g$, one has to distinguish the density fluctuations which are correlated with the enthalpy from those which are not. Even a pure shear rearrangement which increases the structural energy increases the volume because of the anharmonicity of the interatomic potential. This second kind of density fluctuations is intimately related to the additional thermal expansion of the undercooled liquid.

A quantitative thermodynamic treatment of the additional thermal expansion has been given in the parallel paper \cite{pri}. According to this treatment, one can determine the coupling constant ratio $\delta B/\delta G$ from the measurements at $T_g$ via
\begin{equation}\label{dbdg}
	\frac{\delta B}{\delta G}=\frac{B\Delta\kappa(\Pi-1)}{1+B\Delta\kappa(\Pi-1)}\frac{B}{G},
\end{equation}
where $\Pi$ is the Prigogine-Defay ratio of the glass transition \cite{jackle}
\begin{equation}\label{prigo}
	\Pi=\frac{\Delta c_p\Delta\kappa}{(\Delta\alpha)^2T_g}=\frac{\overline{\Delta H^2}\ \ \overline{\Delta V^2}}{(\overline{\Delta H\Delta V})^2},
\end{equation}
which relates the increases of the heat capacity at constant pressure per volume unit $\Delta c_p$, of the compressibility $\Delta\kappa$ and of the thermal volume expansion $\Delta\alpha$ at the glass temperature $T_g$ to the additional enthalpy and volume fluctuations $\Delta H$ and $\Delta V$, respectively. If the enthalpy and volume fluctuations are completely correlated, the Prigogine-Defay ratio is one. Eq. (\ref{dbdg}) shows that this is equivalent to $\delta B/\delta G=0$.

\section{Comparison to experiments in glasses and liquids}

The theoretical curve of eq. (\ref{gbgs}) is shown as a continuous line in Fig. 2. Also shown are experimental data from the literature: the full circles are 30 $\gamma_b^2/\gamma_s^2$-values determined for tunneling states in 26 glasses at low temperatures \cite{berret,pohl,bellessa}, the open symbols are measurements of $\Delta B/G=(B-B_0)/G$ in undercooled liquids from Tables I and II.

\begin{figure}[b]
\hspace{-0cm} \vspace{0cm} \epsfig{file=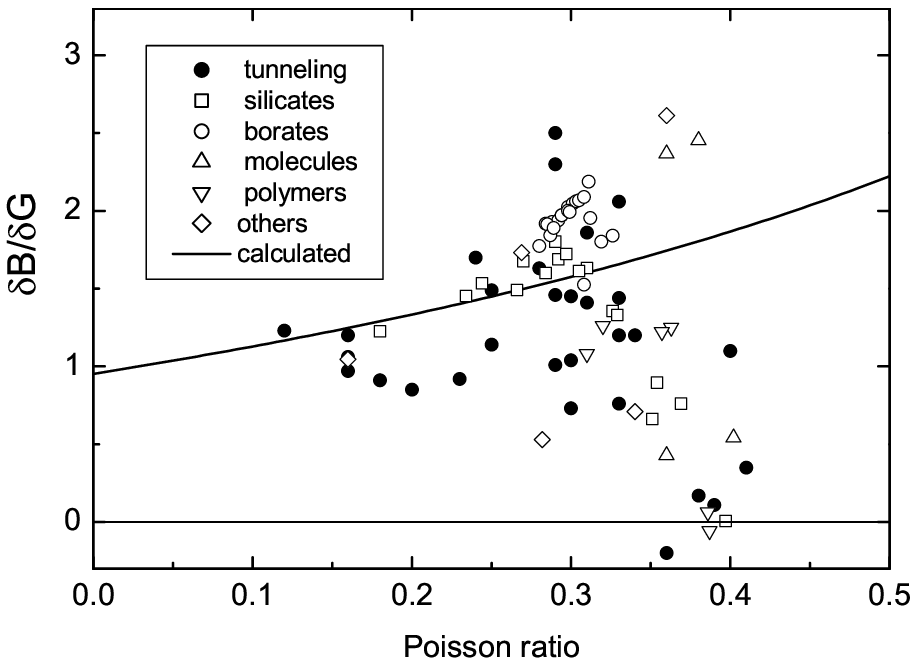,width=7cm,angle=0} \vspace{0cm}\caption{Ratio $\delta B/\delta G$ as a function of the Poisson ratio. The continuous line is the Eshelby prediction of eq. (\ref{dbdg}). The full circles are experimental data for tunneling states in glasses \cite{berret,pohl,bellessa}, the open symbols glass transition values from Tables I and II.}
\end{figure}

\begin{table*}[htbp]
	\centering
		\begin{tabular}{|c|c|c|c|c|c|c|c|c|c|c|c|c|c|}
\hline
subst.&$\overline{M}$&$T_g$&$\rho$&$\alpha_g$&$\alpha_l$&$B$&$B_0$&$c_{pg}$&$c_{pl}$&$\nu$&$\Pi$&\multicolumn{2}{|c|}{$\delta B/\delta G$}          \\
\hline   
 &a. u.& K &kg/m$^3$&\multicolumn{2}{|c|}{10$^{-4}$K$^{-1}$}&\multicolumn{2}{|c|}{GPa}&\multicolumn{2}{|c|}{10$^6$J/m$^3$}& &  &at T$_g$&tunn. \\ 
\hline   
a) molecules: &&&&&&&&&&&&& \\  
glycerol&6.57&183&1316&.90&4.80&11.10&5.56&1.30&2.46&.36&3.73&2.37&\\
glucose&7.50&282&1540&.90&2.60&10.75&6.49&.81&1.30&.38&3.71&2.45&\\
DC704&7.40&214&1080&1.40&4.60&5.26&3.54&1.35&1.65&.40&1.26&.54&\\
o-terphenyl$_{0.67}$o-phenylphenol$_{0.33}$&7.25&236&970&1.69&7.37&5.26&2.93&1.16&1.76&.36&1.19&.43&\\
\hline
b) polymers: &&&&&&&&&&&&& \\  
polyisobutylene&4.66&198&952&1.50&6.20&3.33&2.50&1.20&1.62&.39&.96&-.06&\\
polyvinylacetate&7.16&304&1190&2.80&7.10&3.45&2.00&1.55&2.14&.31&2.22&1.08&\\
polyvinylchloride&10.41&350&1370&2.00&5.70&4.17&2.27&1.59&2.00&.36&1.72&1.25&\\
polystyrene&6.50&362&1031&2.30&5.73&3.12&1.64&1.55&1.80&.36&1.70&1.22&\\
polymethylmethacrylate&6.66&378&1149&2.57&6.06&3.33&1.72&1.75&2.10&.32&2.14&1.26&1.27$^a$\\
bisphenol-A-polycarbonate&6.91&423&1159&2.55&5.99&2.93&1.88&1.45&1.72&.39&1.03&.06&0.11$^b$\\
\hline
c) others: &&&&&&&&&&&&& \\  
GeO$_2$&34.86&933&3590&.27&.76&23.87&8.08&2.42&2.61&.16&6.85&1.05&0.9$^b$\\
Ca$_2$K$_3$(NO$_3$)$_7$&19.14&340&2174&1.20&3.50&16.66&7.69&2.07&3.24&.36&4.57&2.61&\\
selenium&78.96&304&4167&1.70&4.20&4.16&3.33&1.38&2.13&.34&2.37&.71&0.8$^c$\\
anorthite&21.39&1093&2699&.19&.52&40.50&20.60&3.02&3.97&.27&19.04&1.73&\\
diopside&21.70&993&2861&.35&1.25&32.20&24.10&3.05&4.67&.28&2.10&0.53&\\
\hline		
		\end{tabular}
	\caption{Coupling ratio $\delta B/\delta G$ from Prigogine-Defay data and Poisson ratios for molecules, polymers and other glass formers at the glass transition. DC704 is a diffusion pump oil,  tetramethyl-tetraphenyl-trisiloxane. References for Prigogine-Defay data: anorthite and diopside Dingwell et al \cite{dingwell} and Schilling et al \cite{schilling}; polymethylmethacrylate Sane and Knauss \cite{sane}, Kr\"uger et al \cite{krueger}, Bares and Wunderlich \cite{bares} and Schwarzl \cite{schwarzl}; all other substances from the supplement of Gundermann et al \cite{gundermann}. References for Poisson ratios (or for the shear modulus $G$, because $B$ is known from the Prigogine-Defay data):  glycerol Fioretto et al \cite{fioretto}; glucose Meyer and Ferry \cite{meyer}; DC704 Niss et al \cite{niss}; o-terphenyl$_{0.67}$o-phenylphenol$_{0.33}$ assumed to be the one of OTP from T\"olle et al \cite{toelle}; polyisobutylene Litovitz and Davis \cite{litodavis}; polyvinylacetate Donth et al \cite{donthp}; polyvinylchloride Kono \cite{kono}; polystyrene Takagi et al \cite{takagi}; polymethylmethacrylate Kr\"uger et al \cite{krueger}; bisphenol-A-polycarbonate Patterson \cite{patterson}; GeO$_2$ Ananev et al \cite{ananev}; Ca$_2$K$_3$(NO$_3$)$_7$ Torell and Aronsson \cite{torell}; selenium Soga et al \cite{soga}; anorthite and diopside Schilling et al \cite{schilling}. Tunneling data $^a$Berret and Meissner \cite{berret}; $^b$Pohl et al \cite{pohl}; $^c$Bellessa \cite{bellessa}.}
	\label{tab:rrse}
\end{table*}

The experimental values for the tunneling states were taken from the data collections of Berret and Meissner \cite{berret}, Pohl, Liu and Thompson \cite{pohl} and Bellessa \cite{bellessa}. Berret and Meissner report values for the tunneling state coupling constants $\gamma_l$ and $\gamma_t$ to longitudinal and transverse sound waves, respectively. $\gamma_t$ agrees with the $\gamma_s$ of the present paper. For $\gamma_l$, one has the relation
\begin{equation}\label{gammal}
	\gamma_l^2=\gamma_b^2+\frac{4}{3}\gamma_s^2,
\end{equation}
because the longitudinal modulus $M=B+4G/3$. Thus one can calculate $\delta B/\delta G=\gamma_b^2/\gamma_s^2$ from $\gamma_l$ and $\gamma_t$. Pohl, Liu and Thompson \cite{pohl} report values of $C_l$ and $C_t$ from tunneling plateaus in the damping of longitudinal and transverse waves at low temperatures in glasses. From these and the sound velocities ${\rm v}_l$ and ${\rm v}_t$, one can again calculate $\gamma_l^2/\gamma_t^2$ via
\begin{equation}
\frac{\gamma_l^2}{\gamma_t^2}=\frac{C_l{\rm v}_l^2}{C_t{\rm v}_t^2}.
\end{equation}
Bellessa \cite{bellessa} reports values of $n_0\gamma_l^2$ and $n_0\gamma_t^2$ from the temperature dependence of the sound velocities at low temperature in selenium and in three metallic glasses, where $n_0$ is the density of tunneling states. From the ratio of these two values, one can again calculate $\delta B/\delta G$ as from the data of Berret and Meissner. 

The error is large; if $\gamma_l^2$ and $\gamma_t^2$ are measured with an error of 10 \%, the error of $\delta B/\delta G$ is 40 \%. But the scatter of the data is even larger than the one expected from the error. This is a first indication for a large substance dependence of $\delta B/\delta G$.

Table I and Table II compile literature data for the Prigogine-Defay ratio and the Poisson ratio. With these, one can calculate $\delta B/\delta G$ from eq. (\ref{dbdg}). The results in Fig. 2 show a similar scatter and a similar dependence on the Poisson ratio as those determined from the tunneling states at low temperatures. Not only the glass transition data, but also the tunneling state coupling constants follow the theoretical curve of eq. (\ref{gbgs}) at small Poisson ratio $\nu$. At about $\nu=1/3$, the value for central forces in close packing, the values begin to decline away from the theoretical curve, reaching zero at about $\nu=0.4$.

The equality of the coupling ratios at the glass transition and for the low temperature tunneling states is the first main result of the present paper. Note that the few substances in Table I and II where both couplings have been measured show agreement within experimental error.

The second main result is the gradual breakdown of the theoretical prediction as the Poisson ratio begins to approach higher values. Obviously, the assumption of a constant density of stable structural states in distortion space, on which the theoretical prediction is based, breaks down as the Poisson ratio of the glass approaches the value 1/2 of the liquid. The reason for this breakdown is not yet clear, but it tends to occur in close-packed systems \cite{pri}. 

\section{Summary}

The Eshelby concept of a local structural rearrangement within a surrounding elastic matrix allows to derive the coupling ratio $\delta B/\delta G$ of structural relaxation processes to an external compression and an external shear, respectively. Assuming a constant density of stable structures in distortion space, one finds a ratio between one and two, increasing with increasing Poisson ratio.

The comparison to literature data from the low temperature tunneling states in glasses provides essentially the same answer as the comparison to literature glass transition data: At low Poisson ratio, the theoretical prediction is obeyed, though with a large scatter from a strong substance dependence of the ratio. Above the Poisson ratio 1/3, the coupling ratio decreases, reaching zero at a Poisson ratio of about 0.4. At this point, the Prigogine-Defay ratio of the glass transition reaches unity, implying a perfect correlation of the additional enthalpy and the density fluctuations which distinguish the liquid from the glass.

Helpful discussions with Reiner Zorn, Herbert Schober, Miguel Angel Ramos, Michael Ohl and Andreas Wischnewski are gratefully acknowledged.

\end{document}